\documentstyle[preprint,floats,eqsecnum,tighten,prd,aps]{revtex}
\input epsf.tex
\def\DESepsf(#1 width #2){\epsfxsize=#2 \epsfbox{#1}}
\def\ms{\overline{\rm MS}}
\begin{document}
\preprint{\vbox{
\hbox{INPP-UVA-97-05} 
\hbox{September 1997}} }
\draft

\title{New Relations and Constraints on Quark Spin-flavor Contents\\ 
in\\
Symmetry-breaking Chiral Quark Model}
\author{X. Song}
\address{Institute of Nuclear and Particle Physics\\
Department of Physics, University of Virginia\\
Charlottesville, VA 22901, USA\\
(September 7, 1997)\\
(Revised on November 13, 1997)\\
}

\maketitle

\begin{abstract}
New relations between the quark spin-flavor contents of the 
nucleon and axial weak coupling constants are obtained in 
the chiral quark model with both SU(3) and U(1)-breaking
effects. Using the nonsinglet spin combinations, $\Delta_3$ 
and $\Delta_8$, all spin-flavor observables are functions of 
only one parameter $a$ $-$ probability for the chiral pionic 
fluctuation. The upper and lower bounds of these observables 
are given. The optimum range of $a$, determined by NMC data 
$\bar d-\bar u$, gives a constraint to the cutoff of the 
chiral quark field theory. The model predictions are in good 
agreement with the existing data in this range of $a$. The 
roles of kaon, $\eta$ and $\eta'$ are also discussed.
\end{abstract}




\newpage

\leftline{\bf I. Introduction}

In recent years, two interesting features of the spin-flavor structure
of the nucleon have been revealed. First, the EMC \cite{emc} and later
measurements \cite{slac,slac95,slac97,smc} of polarized structure
functions of the nucleon indicate that the quark spin fraction of the
nucleon is unexpected small. It may be explained by a negative strange 
sea quark polarization, or the anomalous gluon contribution \cite{anom} 
in QCD modified quark parton model. Second, a strong violation of the
Gottfried sum rule (GSR) \cite{gott}, measured by the NMC group
\cite{nmc}, implies that the down-sea content exceeds over the up-sea
($\bar d-\bar u>0$). The flavor asymmetry of the light quark sea 
distributions has also been confirmed in Drell-Yan process \cite{na51},
which gives $\bar u(x)/\bar d(x)\simeq 0.5$ at $x=0.18$. Since the light
quark masses are very small compared with the energy scale in deep 
inelastic processes, the $\bar u/\bar d$ asymmetry cannot be explained 
by the gluon splitting mechanism in the framework of the perturbative 
QCD. Since the gluon is flavorless, the quark sea generated by the gluon
splitting is SU(3) symmetric in the LO QCD, while the NLO evolution only
produces minor violation of SU(3) symmetry of sea. Considering the 
suppression arising from the strange quark mass effects, the SU(3)
symmetry of the sea may be violated, but one expects that the SU(2)
symmetry should hold due to the approximate equality $m_u\simeq m_d$. 
It is obvious that to understand these features, nonperturbative 
mechanism and the mass suppression effects are both needed. Historically, 
the Sullivan process \cite{sulli} shown that the meson cloud of the
nucleon can produce an excess of $\bar d$ over $\bar u$. This mechanism
has been extensively used to explain the deficit of GSR. We only list a
few earlier \cite{meson1} and recent works \cite{meson2,bm96,sbf96}. 
Different models, for instance the Nambu-Jona-Lasinio (NJL) model, 
diquark model, and instanton model etc. were also suggested. A more 
detail list of references can be found in a review paper \cite{kumano97}. 
Among all efforts, the chiral quark model formulated in \cite{mg} seems 
provide a more promising framework to understand the nucleon structure.
This model was first employed by Eichten, Hinchliffe and Quigg \cite{ehq}
to explain both the sea flavor asymmetry and the smallness of the quark
spin fraction. The SU(2) chiral quark model calculation leads to an 
one-parameter fit to the sea flavor asymmetry and the quark spin contents.
However, a full U(3) symmetry calculation produces a flavor symmetric sea. 
To solve this problem, the U(1)-breaking was introduced by Cheng and Li
in \cite{cl}. They obtained a rather good two-parameter fit to the
spin-flavor contents of the nucleon. However, the $0^-$ Goldstone boson
emission with SU(3) symmetry predicts $f_3/f_8=1/3$ and 
$\Delta_3/\Delta_8=5/3$ (the definitions of $\Delta_{3,8}$ and $f_{3,8}$
can be found in (2.1) and (2.8e) in Section II below), which are
inconsistent with the experimental data. To remove this inconsistency, 
the SU(3) breaking effect arising from the suppression of chiral kaonic 
fluctuation was first introduced in \cite{song9603-smw}. A more
generalized version which includes the $\eta$ and $\eta'$ suppression 
effects was subsequently given in \cite{song9605}. A somewhat different 
mass suppression description with $\lambda_8$-breaking was studied in 
\cite{wsk97}. The results shown that the kaonic suppression is important 
not only for removing the inconsistency in SU(3) symmetry description, 
but also for obtaining a better fit to the data. We note that a
generalized version including suppression effects arise from $\eta$ and
$\eta'$ fluctuations was also obtained in \cite{cl1}. The previous works 
need to be improved due to (1) the NA51 data $\bar u(x)/\bar d(x)$ at 
$x=0.18$ was assumed to be equal to the ratio of $\bar u/\bar d$, and 
this is highly questionable, (2) there is no definite rule to determine 
the parameters, and it was not clear which data we should consider first 
in verifying the model predictions. In addition, the previous comparison 
of the model prediction with data, especially in the quark flavor sector, 
also needs to be improved. These issues will be addressed in this paper. 
The main results given in the work \cite{song9605} are briefly reviewed 
in section I. Using the nonsinglet spin combinations $\Delta_3$ and 
$\Delta_8$, the description given in \cite{song9605} is reformed in
section II, and all spin-flavor observables depend only on one parameter. 
In this description, new relations and constraints on quark spin and 
flavor contents are obtained. Numerical results, discussions and brief 
summary are given in sections III.
\bigskip

\leftline{$\bigcirc$~~{\it The chiral quark model}}

The effective chiral quark model describes the nucleon properties in 
the scale range between $\Lambda_{\chi{\rm SB}}$ ($\sim$ 1 GeV) and 
$\Lambda_{\rm QCD}$ ($\sim$ 0.2-0.3 GeV), where the spontaneously 
breaking of chiral symmetry leads to the existence of Goldstone bosons
(GB). The important degrees of freedom are the constituent (dressed)
quarks and Goldstone bosons, and the dominant interaction is the coupling
among the quarks and Goldstone bosons, while the gluon effect is expected
to be rather small. In the chiral fluctuation process, a quark could 
change its spin and flavor by emitting Goldstone bosons. Hence the
spin-flavor contents of the nucleon are determined by the valence quark 
structure and all possible chiral fluctuations. The light quark sea 
asymmetry $\bar u<\bar d$ is attributed to asymmetric $\pi^+$ and $\pi^-$ 
fluctuations, which are originated from the existing flavor asymmetry 
of the valence quark numbers in the proton. The quark spin reduction, 
on the other hand, is due to the spin dilution in the chiral fluctuation
processes $q_{\uparrow}\to q_{\downarrow}+GB$.
\bigskip

\leftline{$\bigcirc$~~\it  SU(3)-breaking from kaonic suppression}

The SU(3)-breaking effect arising from the kaonic suppression was 
studied in \cite{song9603-smw}. The {\it probability} of the chiral
fluctuation 
$u\to d\pi^+$ or $d\to u\pi^-$ is defined by
$$a\equiv |g_8|^2=|\Psi(u\to \pi^+d)|^2=
|\Psi(d\to \pi^-u)|^2
\eqno (1.a)$$
where $g_8$ is the quark-octet meson coupling in the SU(3) symmetry 
description. The U(1) breaking parameter is $\zeta=g_0/g_8\neq 1$, where
$g_0$ denotes the quark-$\eta'$ coupling. The SU(3) breaking parameter 
$\epsilon$ is defined by \cite{song9603-smw}
$$\epsilon=|\Psi(u\to K^+s)|^2/|\Psi(u\to\pi^+d)|^2
\eqno (1.b)$$
which is the ratio between the {\it probabilities} of the chiral 
kaonic and pionic fluctuations. Since the chiral kaon is presumably 
more massive than chiral pions, the amplitude for emitting a kaon 
from a light quark is suppressed and thus $\epsilon< 1$. Considering 
the first order chiral fluctuation, the results were \cite{song9603-smw}
$${\bar u}={a\over 3}(\zeta^2+2\zeta+6)
\eqno (1.2a)$$
$${\bar d}={a\over 3}(\zeta^2+8)
\eqno (1.2b)$$
$${\bar s}={a\over 3}(\zeta^2-2\zeta+10)-3a(1-\epsilon)
\eqno (1.2c)$$
and 
$$\Delta u={4\over 3}-{1\over 9}(8\zeta^2+37)a
+{{4a}\over 3}(1-\epsilon)
\eqno (1.3a)$$
$$\Delta d=-{1\over 3}+{2\over 9}(\zeta^2-1)a
-{{a}\over 3}(1-\epsilon)
\eqno (1.3b)$$
$$\Delta s=-a+a(1-\epsilon)
\eqno (1.3c)$$
For a reasonable value $\epsilon\simeq 0.5$ and the parameters
($a\simeq 0.10$, $\zeta\simeq -1.2$) used in \cite{cl}, we
obtained $f_3/f_8\simeq 0.26$ and $\Delta_3/\Delta_8\simeq 1.94$, 
which are much closer to the data. As mentioned in \cite{song9603-smw}, 
the SU(3) breaking effect arising from the kaonic suppression is 
the key factor to break the SU(3) results $f_3/f_8=1/3$ and
$\Delta_3/\Delta_8=5/3$, and change them in the right direction. 
Taking $\epsilon\to 1$, all $(1-\epsilon)$ terms vanish, thus 
(1.2a-c) and (1.3a-c) reduce into the SU(3) symmetry results given 
in \cite{cl}.
\bigskip

\leftline{$\bigcirc$~~\it  SU(3)-breaking including $\eta$ and
$\eta'$ suppression effects}

Having studied the breaking effect of chiral kaonic suppression, 
generalizing to include the $\eta$ and $\eta'$ suppression is
straightforward and the results were given in \cite{song9605} 
$$u=2+\bar u~, ~~~
{\bar u}={a\over 9}[9+2(3-A)^2+A^2]
\eqno (1.4a)$$
$$d=1+\bar d~, ~~~
{\bar d}={a\over 9}[18+2A^2+(3-A)^2]
\eqno (1.4b)$$
$$s=0+\bar s~, ~~~
{\bar s}={a\over 3}[9+B^2-9(1-\epsilon)]
\eqno (1.4c)$$
and
$$ {{\bar u}\over {\bar d}}=1-{{2A}\over
{(A-1)^2+8}}
\eqno (1.5a)$$
$$ {\bar d}- {\bar u}=a{{2A}\over 3}
\eqno (1.5b)$$
where
$$A\equiv 1-\zeta'+{{1-{\sqrt\epsilon_{\eta}}}\over 2},
\qquad  B\equiv \zeta'-{\sqrt\epsilon_{\eta}}
\eqno (1.6)$$
The parameters $\epsilon_{\eta}$ and $\epsilon_{\eta'}$ 
were defined as
$$|\Psi(u\to \eta(\eta') u)|^2
=|\Psi(d\to \eta(\eta') d)|^2
\equiv\epsilon_{\eta(\eta')} a
\eqno (1.7)$$
where $\zeta'={\sqrt{\epsilon_{\eta'}}}\zeta$. Physically, $\zeta'$
combines both U(1)-breaking (other than mass effect) and mass
suppression effect in the $\eta'$ fluctuation. Hence $\zeta'$ can be
treated as an effective U(1)-breaking parameter. 

The results for the spin contents were \cite{song9605} 
$$\Delta u={4\over 3}-{a\over 9}(8\zeta^{'2}+37)
+{{4a}\over 3}(1-\epsilon)
+{{4a}\over 9}(1-\epsilon_{\eta})
\eqno (1.8a)$$
$$\Delta d=-{1\over 3}+{2a\over 9}(\zeta^{'2}-1)
-{a\over 3}(1-\epsilon)-{a\over 9}(1-\epsilon_{\eta}) 
\eqno (1.8b)$$
$$\Delta s=-a+a(1-\epsilon)
\eqno (1.8c)$$
Taking $\epsilon_{\eta,\eta'}\to 1$ and $\zeta'\rightarrow \zeta$, 
(1.2a$-$c), and (1.3a$-$c) can be easily recovered, where the SU(3) 
breaking effect arises from the kaon suppression only. Assuming 
$\epsilon_{\eta}\simeq\epsilon$ ($m_{\eta}\simeq m_K$) and using data
$\bar d-\bar u$ and a $\bar u/\bar d$ value estimated from some
phenomenological antiquark distributions, a good agreement between the
model predictions and data was obtained \cite{song9605}. 

Several remarks should be made here. (1) The {\it strange quark 
polarization} $\Delta s$ is {\it not} affected by introducing the 
suppression of $\eta$ and $\eta'$ mesons, because the $\eta$ and 
$\eta'$ are all strangeness-0 and spin-0 mesons. They provide equal
components of $s_{\uparrow}$ and $s_{\downarrow}$ and give vanishing 
contribution to the strange spin content. Only nonvanishing 
contribution to $\Delta s$ is coming from the kaons, and it decreases 
the strange flavor content $\bar s$ by $3a(1-\epsilon)$, and reduce 
the $magnitude$ of the strange quark polarization $\Delta s$ by 
$a(1-\epsilon)$ as given in \cite{song9603-smw}. However, the suppression 
effects of chiral $\eta$ and $\eta'$ fluctuations do give some
contributions to nonstrange quark polarizations: $\Delta u$ becomes 
more positive and $\Delta d$ is more negative. This certainly
gives a better fit to the data. (2) The special combinations $A$ 
and $B$, appeared in quark flavor contents (1.4a-c), {\it do not} appear
in (1.8a-c). This is because the $\eta$ and $\eta'$ are spin-0 neutral 
mesons, the $A$, $B$ terms appeared in the spin-up quark content
($q_{\uparrow}$) are completely the same as those in the spin-down 
case ($q_{\downarrow}$), hence they cancel each other in $\Delta q$.
(3) All antiquarks in the sea are unpolarized
$$\Delta\bar u=0~,~~~~\Delta\bar d=0~,~~~~\Delta\bar s=0~,
\eqno (1.8d)$$
which hold for all chiral quark models with only first order Goldstone
boson fluctuations, because the Goldstone bosons consist of equal 
components of $\bar q_{\uparrow}$ and $\bar q_{\downarrow}$ 
This prediction is consistent with recent semi-inclusive data 
\cite{smc96} (see Table III), and has been used to explain the 
baryon magnetic moments \cite{cl2}.

Since (1.8c) provides a simple relation between $\Delta s$ and the 
parameters $a$ and $\epsilon$, it is tempting to use (1.8c) and 
the data $\Delta s$ to determine $\epsilon$, if one has obtained $A$ 
and $a$ from (1.5a-b) by using data $\bar u/\bar d$ and $\bar d-\bar u$.
However, the data given by NA51 experiment \cite{na51} only provides 
{\it one value} of the ratio ${\bar u}(x)/{\bar d}(x)$ at $x\simeq 0.18$, 
while we need a ratio of the overall integrated quantities $\bar u$ and 
$\bar d$, which is almost undetermined. In addition, $\Delta s$ measured 
in the deep inelastic polarized lepton nucleon scattering is less 
precisely known. In Section II, a new approach and an one-parameter
description will be introduced. 
\bigskip

\leftline{\bf II.~ New relations and constraints}

Let us reform the formalism (1.8a-c) and (1.4a-c), and rewrite the
observables of spin-flavor contents as functions of one parameter $a$
only. Defining the nonsinglet spin combinations, or nonsinglet axial
charges 
$$\Delta_3\equiv\Delta u-\Delta d~,~~~~ 
\Delta_8\equiv\Delta u+\Delta d-2\Delta s
\eqno (2.1)$$
and assuming $\epsilon_{\eta}=\epsilon$, from (1.8a$-$c), one 
obtains
$$\zeta'^2={1\over {2a}}[3(1-\Delta_8)-{\Delta}]-{7\over 2}
\eqno (2.2a)$$
$$\epsilon=1-{{\Delta}\over {2a}}
\eqno (2.2b)$$
where
$$\Delta\equiv {3\over 5}\Delta_3-\Delta_8
\eqno (2.3) $$
In the SU(6) valence quark model, $\Delta u=4/3$, $\Delta d=-1/3$, 
$\Delta s$=0, and $\Delta_3$=5/3,$\Delta_8=1$, which lead to 
$\Delta=0$. Hence $\Delta\neq 0$ measures the deviation from the 
naive quark model prediction. According to Cabibbo's description, 
the nonsinglet axial charges $\Delta_3$ and $\Delta_8$, related to 
the weak axial couplings, can be measured in the hyperon 
$\beta$-decays \cite{cr93,skm96,rat96,pdg96},  
$$\Delta_3=\Delta u-\Delta d=({{G_A}\over {G_V}})_{n\to p}=F+D
\eqno (2.4a) $$
$$\Delta_8=\Delta u+\Delta d-2\Delta s=3F-D
\eqno (2.4b) $$
Experimentally, they are rather precisely known 
$$\Delta_3=1.2573\pm 0.0028~,~~~~\Delta_8=0.579\pm 0.025
\eqno (2.5) $$
hence 
$$\Delta=0.175\pm 0.025
\eqno (2.6) $$
which is about $15-20\%$ deviation from zero. Having known $\Delta_3$,
$\Delta_8$ and $\Delta$, from (2.2a-b) and (1.6), all spin and flavor
contents are now functions of {\it one parameter $a$ only}.
\bigskip

\leftline{$\bigcirc${~~\it New relations on quark spin contents}}

Using (2.2a-b), the quark spin contents (1.8a-c) can be rewritten as
$$\Delta u={4\over 5}\Delta_3-a
\eqno (2.7a)$$
$$\Delta d=-{1\over 5}\Delta_3-a
\eqno (2.7b)$$
$$\Delta s={{\Delta}\over 2}-a
\eqno (2.7c)$$
and the total quark spin is 
$$\Delta\Sigma=\Delta u+\Delta d+\Delta s={3\over 5}\Delta_3+
{{\Delta}\over 2}-3a
\eqno (2.7d)$$
Equations (2.7a-d), derived from the chiral quark model with both 
SU(3) and axial U(1) breakings, provide new and simple relations 
connecting the quark spin contents and weak axial couplings. 
The comparison with the Skyrme model \cite{bek88} and naive quark 
model results is shown in Table 1. 
\bigskip

\leftline{$\bigcirc${~~\it Observables related to the quark flavor
contents}}

(a) The ratio of total antiquark contents to total quark contents
is 
$$r_{{\bar q}/q}\equiv {{\sum\bar q}\over {\sum q}}={1\over 2}(
1-{1\over {K_1+1+3a}})~,~~~~K_1\equiv 1-\Delta_8-3\Delta/2
\eqno (2.8a)$$

(b) The ratio of the total strange sea to the light antiquark contents is
$$ r_{{2\bar s}/\bar u+\bar d}\equiv {{2\bar s}\over {\bar u+\bar d}}
={{B^2+9(1-{\Delta}/(2a))}\over {A^2-3A+9}}
\eqno (2.8b)$$

(c) The ratio of the total strange sea to the light quark contents is
$$ r_{{2\bar s}/u+d}\equiv {{2\bar s}\over {u+d}}
=2a{{B^2+9(1-{\Delta}/(2a))}\over {9+2a(A^2-3A+9)}}
\eqno (2.8c)$$

(d) The ratio of the total strange sea to total quark and antiquark
contents is
$$f_s\equiv{{s+\bar s}\over {\sum(q+\bar q)}}={{2a}\over
9}{B^2+9(1-\Delta/(2a))\over {K_1+1+3a}}
\eqno (2.8d)$$

(e) Defining $f_q\equiv(q+\bar q)/\sum(q+\bar q)$,
$f_3\equiv f_u-f_d$ and $f_8\equiv f_u+f_d-2f_s$, one has
$${{f_3}\over {f_8}}\equiv{{u-d+\bar u-\bar d}\over {u+d-2s+\bar u+\bar
d-2\bar s}}={1\over 3}(1+K_2)^{-1}
\eqno (2.8e)$$
with $K_2\equiv 4a[(A^2-B^2)+9\Delta/(2a)]/(1-4aA/3)$, where $A$, and $B$
are functions of $a$ determined from (1.6), and (2.2a-b). Taking 
$\epsilon,\epsilon_{\eta}\to 1$, then $\Delta\to 0$, $A\to -B$, one has
$K_2\to 0$ and $f_3/f_8\to 1/3$. 

(f) There is a relation among the observables given in (2.8b-d) 
$$f_s=(1+{1\over {r_{2\bar s/u+d}}}+
{1\over {r_{2\bar s/\bar u+\bar d}}})^{-1}.
\eqno (2.8f)$$
To obtain the upper and lower bounds of the spin and flavor 
observables, we need to discuss the {\it physically allowed range} 
of $a$ and the $sign$ of $\zeta'$. 
\bigskip

\leftline{$\bigcirc${~~\it Parameter a}}

Since $\zeta'^2$ and $\epsilon$ must be positive, from (2.2a-b) one
obtains the lower and upper bounds of $a$
$$a_{min}={{\Delta}\over 2}~,~~~~~~~
a_{max}={1\over 7}[3(1-\Delta_8)-\Delta]
\eqno (2.9a)$$
Using (2.5) and (2.6), we have $a_{min}=0.088\pm 0.012$ and
$a_{max}=0.155\pm 0.004$, hence the allowed range of $a$ is 
approximately
$$0.08~~<~~a~~<~~0.16
\eqno (2.9b)$$
In section III, we will show that this range can be further narrowed 
by using NMC data $\bar d-\bar u$. One interesting prediction, from 
(2.7c), is that the {\it upper bound of the strength of the strange 
quark polarization} is 
$$|\Delta s|_{max}=a_{max}-{{\Delta}\over 2}\simeq
0.08
\eqno (2.10)$$

The probability $a$ for an up-quark splitting into a down-quark and a
$\pi^+$ can be estimated in the chiral field theory \cite{ehq}
$$a={{g_A^2m^2}\over {8{\pi}^2f_{\pi}^2}}\int_0^1zdz\Theta(z_{max}-z)
\{ {\rm ln}[ {{\Lambda^2+m_{\pi}^2}\over
{\tau(z)+m_{\pi}^2}}] +m_{\pi}^2
[{1\over {\Lambda^2+m_{\pi}^2}}
-{1\over {\tau(z)+m_{\pi}^2}}]\}
\eqno (2.11)$$
where $g_A\simeq 0.75$ is dimensionless axial-vector coupling, 
$f_{\pi}\simeq 0.093$ GeV the pion decay constant, $m\simeq 0.35$ GeV 
the constituent mass of $u$ or $d$-quarks, $m_{\pi}$ the pion mass,
$\Lambda$ the ultraviolet cutoff, $\tau(z)=m_u^2z^2/(1-z)$ and 
$z_{max}={{\Lambda^2}\over {2m^2}}(\sqrt{1+{{4m^2}\over {\Lambda^2}}}-1)$.
The range given in (2.9b) can be well reproduced by taking $\Lambda\simeq
1.20-2.57$ GeV.
\bigskip

\leftline{$\bigcirc${~~\it Sign of $\zeta'$}}

From (1.5b), (1.6), and (2.2a-b), one obtains $\zeta'$ as function 
of $a$
$$\zeta'=1-{{3\delta}\over {2a}}+
{1\over 2}[1-{\sqrt{1-{{\Delta}\over {2a}}}}]
\eqno (2.12)$$
where $\delta\equiv\bar d-\bar u$ is {\it input}. we plot the 
$\zeta'-a$ curves for $\delta=0.130$,
0.147, and 0.164, in Fig. 1. Here and in what follows we literally 
use the central value of NMC result $\bar d-\bar u=0.147$ and restrict
the error to $\pm 0.017$ in the numerical evaluation (see discussion 
of possible large error on (3.1) in section III). For the allowed 
range of $a$ given in (2.9b), the U(1)-breaking parameter $\zeta'$ 
is $negative$. 
\bigskip

\leftline{$\bigcirc$~~{\it Range of $\epsilon$ }}

From (2.2b), one has
$$1-{{\Delta}\over {2a_{min}}}~<~\epsilon~<~1-{{\Delta}\over {2a_{max}}}
\eqno (2.13a)$$
or
$$0~<~\epsilon~<~1-{{a_{min}}\over {a_{max}}}
\eqno (2.13b)$$
Hence the upper bound of $\epsilon$ is approximately 0.43.
\bigskip

\leftline{$\bigcirc$~~{\it Ranges of $\Delta u$, $|\Delta d|$, 
$|\Delta s|$ and $\Delta\Sigma$ }}

From (2.7a-d), we obtain
$${4\over 5}\Delta_3-a_{max}~<~\Delta u~<~{4\over
5}\Delta_3-a_{min}
\eqno (2.14a)$$
$${1\over 5}\Delta_3+a_{min}~<~|\Delta d|~<~{1\over
5}\Delta_3+a_{max}
\eqno (2.14b)$$
$$-{{\Delta}\over 2}+a_{min}~<~|\Delta s|~<~-{{\Delta}\over 2}+a_{max}
\eqno (2.14c)$$
and
$${3\over 5}\Delta_3+{{\Delta}\over 2}-3a_{max}~<~
\Delta\Sigma~<~{3\over 5}\Delta_3+
{{\Delta}\over 2}-3a_{min}
\eqno (2.14d)$$
where $a_{max}$ and $a_{min}$ are given in (2.9a). Similarly
we can obtain the constraints for other interesting observables.
We note that (2.14c) gives a range of {\it negative} strange sea
polarization
$$-0.08<\Delta s<0
\eqno (2.14e)$$

\bigskip

\leftline{\bf III~ Numerical results and discussion}

As mentioned above, a narrowed range of $a$ can be obtained by using 
the NMC data. Substituting (2.2a-b) into expression $A$ in (1.6), then
from (1.5b), $\bar d-\bar u$ can be written as function of $a$, which 
is plotted in Fig.2. The straight lines $\delta\equiv\bar d-\bar u=0.130$,
0.147, and 0.164 are also shown. They lead to $a=0.144$, 0.153, and 0.155, 
thus we obtain a range of parameter $a$ determined by the NMC data 
$$0.144~<~a~<~0.155
\eqno (3.1)$$
which, we call {\it the optimum range of $a$}, is much narrower than
that given in (2.9b). If we take $\bar d-\bar u=0.147\pm 0.024$, the
lower value of $a$ would be 0.138, and the upper value $a=0.155$ is
still the same, because in our formalism, this is physically allowed
upper bound of $a$ determined by (2.2a). From the range (3.1), one 
obtains
$$0.39~<~\epsilon~<~0.43
\eqno (3.2)$$
The range (3.1) implies that the probability of the lowest order chiral
pion fluctuation (for example $u\to d+\pi^+$) is of the order $15\%$,
and the range (3.2) implies that the probability ($\epsilon a$) of the
chiral kaon fluctuation (for example $u\to s+K^+$) is of the order $6\%$. 
Under the approximation $\epsilon_{\eta}\simeq \epsilon$, 
the probability of the chiral $\eta$ fluctuation (for instance, 
$u\to u+\eta$ etc) is also of the order $6\%$. They give 
restrictions to the calculation from any model wave functions.  
Similarly we can obtain the ranges for other observables. For 
instance, in the range (3.1), the predicted quark spin contents 
are $0.85\leq\Delta u\leq0.86$, $-0.41\leq\Delta d\leq -0.40$, 
and $-0.07\leq\Delta s\leq -0.06$.  These and other model
predictions are listed in Table II and Table III. For comparison, 
the SU(3) symmetry case \cite{cl} and SU(3) breaking results 
assuming the kaon suppression only \cite{song9603-smw} are also listed.  
\bigskip

\leftline{$\bigcirc$~~\it Quark spin contents}

To illustrate the $a$-dependence of the quark spin contents, we plot
$\Delta u$, $-\Delta d$ and $-\Delta s$ as functions of $a$ in Fig. 3. 
It can be seen that the model prediction agrees very well with most 
recent DIS data \cite{smc97,ek} in the range (3.1), see also Table III. 
Two remarks should be made here. First, the data of $\Delta u$, $\Delta
d$, and $\Delta s$ listed in Table III are actually $a_u$, $a_d$, and 
$a_s$ $-$ the proton matrix elements of quark axial vector currents,
which are obtained from the nonsinglet axial charges $a_{3,8}$,
i.e. our $\Delta_{3,8}$, and singlet axial charge $a_0$ defined by 
$a_3=a_u-a_d$, $a_8=a_u+a_d-2a_s$, and $a_0=a_u+a_d+a_s$.
The axial charges $a_{3,8}$ and $a_0$ are determined by using the
hyperon $\beta$-decay data (2.5) and the first moment of $g_1(x,Q^2)$ 
(see (3.8) below) measured from polarized DIS experiments. In general, 
the proton matrix element of singlet axial current $a_0$ is sum of 
quark spin contribution $\Delta\Sigma$ and a gluonic term. 
However, the separation of terms proportional to $\Delta\Sigma$ 
and $\Delta G$ is $arbitrary$ and depends upon the {\it factorization
scheme}. In the Adler-Bardeen (AB) scheme, the chirality is preserved 
and transitions between quarks of different helicities are forbidden 
to any order in perturbation theory, and
$a_0(Q^2)=\Delta\Sigma-n_f{{\alpha_s(Q^2)}\over {2\pi}}\Delta G(Q^2)$, 
where $\Delta\Sigma$ is {\it independent} of $Q^2$, $n_f$ is the number
of active quark flavors, and $\Delta G(Q^2)$ is gluon polarization. 
For the contributions from individual quark flavors, one has 
$a_q(Q^2)=\Delta q-{{\alpha_s(Q^2)}\over {2\pi}}\Delta G(Q^2)$, 
($q=u,d,s$ in this paper), where $\Delta q$'s are also {\it independent}
of $Q^2$. These are the standard results of axial anomaly \cite{anom}. 
On the contrast, in the $\ms$ (or Gauge-invariant) scheme, 
the chirality is not conserved and one has $a_0(Q^2)=\Delta\Sigma(Q^2)$,
and $a_q(Q^2)=\Delta q(Q^2)$, where $\Delta\Sigma(Q^2)$ and 
$\Delta q(Q^2)$'s are now {\it dependent} of $Q^2$. The anomalous 
gluon contribution has been absorbed into $\Delta\Sigma$ or $\Delta q$ 
as an effective sea contribution. We do not discuss some disputes of 
which scheme is more appropriate (for instance, see \cite{cheng96}), 
but note that $a_0(Q^2)$ and $a_q(Q^2)$'s are {\it scheme-independent}
quantities. In Table III, 
the chiral quark model predictions are compared to the $a_q(Q^2)$'s 
rather than the scheme-dependent $\Delta q$'s. Second, we note that 
in the AB scheme, the gluonic term is independent of $Q^2$ at LO, 
because the ln$Q^2$ growth of $\Delta G(Q^2)$ is compensated by 
the 1/ln($Q^2$) decrease of running coupling $\alpha_s^{({\rm LO})}(Q^2)$ 
at LO, and the singlet axial charge $a_0(Q^2)$ is independent of 
$Q^2$ at the same order (for instance see \cite{song89}). Detail
analysis shows that $a_0$ decreases very slowly with $Q^2$ at NLO 
\cite{kod}. Hence as far as we assume that the chiral quark model 
results of $\Delta u$, $\Delta d$ and $\Delta s$ can be identified 
as the corresponding DIS observables $a_u$, $a_d$, and $a_s$ at the 
range (0.2 GeV)$^2<Q^2<$(1.0 GeV)$^2$, and the perturbative QCD can be 
used down to this low $Q^2$ scale, the comparison of the chiral 
quark model prediction with the deep inelastic polarized data is
meaningful. However, although the perturbative QCD 
evolution approach has been successfully used down to $Q^2\simeq$ 
0.23 GeV$^2$ in \cite{grv95}, it is not clear if the approach still 
holds below this $Q^2$. We also note that the DIS data 
$\Gamma_1^{p,n,d}(Q^2)$ listed in Table III do not depend on the 
factorization scheme.
\bigskip

\leftline{$\bigcirc$~~{\it $\bar u/\bar d$, $2\bar s/(\bar u+\bar d)$, 
$2\bar s/(u+d)$, and $\sum\bar q/\sum q$}}

We plot $\bar u/\bar d$, $2\bar s/(\bar u+\bar d)$, and $2\bar s/(u+d)$
in Fig. 4. The CCFR data \cite{ccfr95} are also shown. It should be 
noted that the CCFR data only give the ratios of the strange quark 
momentum to light quark or light antiquark momentum, or total antiquark
momentum to total quark momentum. Defining $Q\equiv\int_0^1xq(x)dx$, 
the data show 
$$\kappa\equiv{{2S}\over {\bar U+\bar D}}=0.477\pm 0.051
\eqno (3.3)$$
$$\eta\equiv{{2S}\over {U+D}}=0.099\pm 0.008\pm 0.004
\eqno (3.4)$$
$${{\sum\bar Q}\over {\sum Q}}=0.245\pm 0.005~
\eqno (3.5)$$
It should be noted that in our notation, $q\equiv\int_0^1q(x)dx$.  
If the integral $\int_0^1q(x)dx$ is $finite$, one has $Q=x_0^{(q)}q$, 
where the mean value theorem is used and $x_0^{(q)}$ is between 0 and 1.
The $x_0^{(q)}$ value depends on the shape of function $q(x)$. Comparing 
to the quark distributions, the antiquark distributions are dominate 
in the smaller $x$ region, hence $x_0^{(\bar q)}<x_0^{(q)}$. For similar
reason, $x_0^{(\bar s)}<x_0^{(\bar u+\bar d)}$. Hence we expect
$${{2S}\over {U+D}}~<~{{2s}\over {u+d}}~,~~~~~
{{S}\over {\bar U+\bar D}}~<~{{s}\over {\bar u+\bar d}}~,~~~~~
{{\sum\bar Q}\over {\sum Q}}~<~{{\sum\bar q}\over {\sum q}}
\eqno (3.6)$$
It can be seen, from Fig.4, the model predictions of
$2\bar s/(\bar u+\bar d)$, and $2\bar s/(u+d)$, and corresponding data
satisfy (3.6) very well. For the ratio $\sum\bar q/\sum q$, (3.6) seems
not hold, but the model prediction $\sum\bar q/\sum q=0.238$ is very 
close to the data (3.5).

For the ratio $\bar u/\bar d$, as Field and Feynman suggested two 
decades ago \cite{ff77}, $\bar d(x)$ and $\bar u(x)$ have different 
large $x$ behavior, and $\bar u/\bar d$ may differs from 
$\bar u(x)/\bar d(x)$. To show this, we need to know the shape of 
the light antiquark sea distributions, which have not been well 
determined. There are several phenomenological distributions from 
fitting the DIS data in the literature (see, for instance, 
\cite{mrs95,mrs96,cteq97}). Since we do not intend to present a model 
calculation on the quark distributions in this paper, we simply take
$x(\bar u(x,\mu^2)+\bar d(x,\mu^2))$ and 
$x(\bar u(x,\mu^2)-\bar d(x,\mu^2))$ at $\mu^2$=0.34 GeV$^2$ 
from \cite{grv95} to represent the chiral quark model distributions, 
which gives $\bar u/\bar d\simeq 0.83$ and 
$[\bar u(x)/\bar d(x)]_{x=0.18}\simeq 0.46$ at $Q^2=0.34$ (GeV)$^2$. 
After the QCD evolution, the ratio $\bar u(x)/\bar d(x)$ increases
and is approximately 0.53 at $Q^2=4$ GeV$^2$. Hence the chiral quark
model prediction $\bar u/\bar d\simeq 0.65$ is not necessarily to
contradict with the NA51 result $[\bar u(x)/\bar d(x)]_{x=0.18}=0.51
\pm 0.06$. However, a more detail quantitative study on the difference
between these two ratios is needed. 
\bigskip

\leftline{$\bigcirc$~~{\it $f_s$ and $f_3/f_8$}}

For the ratio $f_s$, the model prediction agrees very well 
with the phenomenological values (see Table II). In Table II, 
$f_s=0.10\pm 0.06$ is taken from \cite{gls91}, where $f_s=y/(2+y)$ 
and $y=1-\sigma_0/\sigma$, with $\sigma_0=35\pm 5$ MeV, and 
$\sigma\simeq 45$ MeV. If one takes $\sigma_0=25\pm 5$ 
MeV, then $f_s=0.18\pm 0.03$, which was used in the previous works
\cite{cl,song9603-smw,song9605,wsk97,cl1}. The value
$(f_s)_{lattice}=0.15\pm 0.03$,
listed in Table II, is taken from the lattice QCD calculation 
\cite{dll96}. 
From (3.3) and (3.4), one has $1/\kappa+1/\eta
=(U+\bar U+D+\bar D)/(2S)=(U+\bar U+D+\bar D)/(S+\bar S)$, 
where the asumption $\bar S=S$
(i.e. $\int_0^1x{\bar s}(x)dx=\int_0^1x{s}(x)dx$) has been used. 
Considering $\sum\limits(Q+\bar Q)=U+\bar U+D+\bar D+S+\bar S$, 
we have 
$${{2S}\over {\sum(Q+\bar Q)}}=(1+{1\over {\kappa}}+{1\over {\eta}})^{-1}
\eqno (3.7)$$
Using the CCFR data, one obtains $0.076\pm 0.022$ for the r.h.s. of (3.7).
This number has been used in Table II. Hence our
prediction $f_s=0.10$ satisfies $f_s=2s/\sum(q+\bar q)>2S/\sum(Q+\bar Q)$.
For the ratio $f_3/f_8$, the model prediction is consistent with the 
phenomenological value $f_3/f_8=0.23\pm 0.05$ \cite{cl}. 
\bigskip

\leftline{$\bigcirc$~~{\it First moments of $g_1^{p,n,d}$} }

The first moment of $g_1^{p}$ can be expressed in terms of the 
axial charges $a_{3,8}$ and $a_0$
$$\Gamma_1^p(Q^2)={{C_{\rm NS}(Q^2)}\over {12}}(a_3+{{a_8}\over 3})
+{{C_{\rm S}(Q^2)}\over {9}}a_0(Q^2)
\eqno (3.8)$$
where $C_{\rm NS}(Q^2)$ and $C_{\rm S}(Q^2)$ are the QCD radiative
coefficient functions \cite{larin94}, which depend on the strong 
coupling $\alpha_s(Q^2)$. Similar expression for the neutron can be
obtained by changing $a_3$ to $-a_3$. For the deuteron, we use
$\Gamma_1^d=\eta(\Gamma_1^p+\Gamma_1^n)$, where $\eta=0.4565$.
The results are given in Table III. For SLAC data, we use 
$\alpha_s$(5 GeV$^2)\simeq 0.30$ for $\Gamma_1^{n}$ \cite{slac97}, 
and $\alpha_s$(3 GeV$^2)\simeq 0.35$ for $\Gamma_1^{p,d}$ \cite{slac95}.
For SMC data at $Q^2=10$ GeV$^2$, we use $\alpha_s$(10 GeV$^2)\simeq
0.25$. It can be seen from Table III that the chiral quark model
prediction agrees very well with the existing data. 
\bigskip

Having shown a remarkable success of the model, some comments are 
in order.

(1) Using the nonsinglet axial charges, which can be determined by
accurate low energy hyperon $\beta$-decay data, the symmetry breaking 
chiral quark model results are reformulated and simplified. In this new
description, only one parameter remains. In principle, this parameter 
can be uniquely determined by the accurate data from any
spin-flavor observable, except for $\Delta_3$ and $\Delta_8$. In this
paper, we choose $\bar d-\bar u$ to determine $a$, and find that the 
model with both SU(3) and U(1) breakings provides a very good 
description to almost all existing spin-flavor observables of 
the nucleon.

(2) The breaking effect arising from the suppression of kaonic 
fluctuation is {\it more important} than that from $\eta$ suppression. 
In the limiting case given in \cite{song9603-smw}, $\epsilon_{\eta}\to 1$,
(i.e. 
no $\eta$ suppression, but with U(1)-breaking, $\zeta\neq 1$), the 
overall fit is good and better than the SU(3) symmetry case (see Table 
II and Table III). 

(3) The optimum range of $0.144<a<0.155$ gives a restriction to 
the range of cutoff parameter, $2.30 {\rm GeV}~<~\Lambda~<~2.58 
{\rm GeV}$ in (2.11). However, the range of $\Lambda$ may varies
for different models. For instance, the range of $\Lambda$ used in
\cite{sbf96} is $2.247-5.5$ GeV, which gives $\epsilon\simeq 0.35-0.57$.
This result is quite similar to (3.2).

(4) From the results given in Tables II and III, the optimum fit is
around $a=0.153$, which leads to $\zeta'=-0.233$ and $\epsilon=0.427$. 
It implies that the kaon and $\eta$ fluctuations are strongly suppressed,
and the $\eta'$ suppression, or effective U(1)-breaking, is even stronger.
The relative probabilities of pion, kaon ($\eta$), and $\eta'$
fluctuations are 
$$ \pi~:~K(\sim\eta)~:~\eta'~~\simeq~~ 1~:~0.43~:~0.05
\eqno (3.9)$$
Note that the probability of chiral $\eta'$ fluctuation is $\zeta'^2
a\simeq 0.008$. In the extreme case, $a\to a_{max}$, which leads to
$\zeta'\to 0$ (see (2.2a) and (2.9)). For $a=0.155$, one has 
$\zeta'\simeq -0.09$. Even in this case, $\zeta'^2a\simeq 0.001<<1$, 
the agreement is almost the same as in $a=0.153$ case (see Tables II 
and III). Hence the $\eta'$ plays a minor role (but it means a strong 
U(1)-breaking !), and almost decouples with the quarks in describing 
the spin-flavor structure of the nucleon. This is presumably caused 
by a very heavy mass of $\eta'$. It implies that a better fit to the
existing data requires very small $\eta'$ contribution. This result
is significantly different from that obtained in the original 
U(1)-breaking but SU(3)-symmetry description, where $|\zeta|\simeq 1.2$
leads to an unreasonable large contribution from $\eta'$ fluctuation, 
$\zeta^2a\simeq 0.14-0.15$ (for $a\simeq 0.10$), which is even larger
than the pionic contribution ($\sim a$) !

(5) In obtaining the simple relations (2.7a-c), we have used the
nonsinglet axial charges $\Delta_3$ and $\Delta_8$, and rewrite all
observables into functions of $a$. Since the chiral quark model is
supposed to be a good approximation only at the low scale
(0.2-0.3 GeV)$^2<Q^2<$(1 GeV)$^2$, it is more reasonable to use
$\Delta_3$ and $\Delta_8$, which are related to the low energy 
hyperon $\beta$-decay data, to reform the formalism. Our result shows
that this is a more effective approach to obtain an optimum fit to data.
This approach, however, cannot be applied to the SU(3) symmetry scheme
due to inconsistency. The data $\Delta_3$ and $\Delta_8$ given in (2.5) 
do not satisfy $\Delta_3/\Delta_8=5/3$. Hence, in the SU(3) symmetry case 
we can only choose one of $\Delta_3$ and $\Delta_8$ (we choose $\Delta_3$) 
and another data (we choose $\Delta s=-0.10$ \cite{temp}) as inputs. The 
SU(3) relation $\Delta_8=3\Delta_3/5$ leads to $\Delta_8=0.754$ as shown
in Table III. 

(6) Finally, we note that if one uses $\Delta_3=1.2601\pm 0.0025$ 
\cite{pdg96} in (2.5), then (2.6) becomes $\Delta=0.177\pm 0.025$, 
which does not change our main results (2.7a-d), (2.8a-e), (2.9a), 
(2.12), (2.13a-b) and (2.14a-d), and has only minor impact on the
numerical results.
\bigskip

\leftline{$\bigcirc$~~{\it Summary}}

Using nonsinglet axial charges $\Delta_3$ and $\Delta_8$ , the chiral
quark model results on quark spin-flavor contents are reformulated.
Using the hyperon $\beta$-decay data, the upper and lower bounds
of the spin-flavor observables are given. Special attention is paid
to the comparison of the prediction with data in the quark flavor
sector. We found that the model predictions are 
in good agreement with the existing data in the range $a=0.144-0.155$, 
which gives a constraint to the cutoff in the chiral field theory.
Our result shows that the SU(3) symmetry breaking effects arising 
from the kaon suppression is important, the $\eta$ plays a `fine turning'
role, and the $\eta'$ can even be removed from the description.

\vspace{0.2 cm}

\leftline{\bf Acknowledgements}

This work was supported in part by the U.S. DOE Grant No.
DE-FG02-96ER-40950, the Institute of Nuclear and Particle Physics,
University of Virginia, and the Commonwealth of Virginia.

\bigskip
\bigskip

\vfill\eject

\begin{table}[ht]
\begin{center}
\caption{Quark spin contents of the proton in different models}
\begin{tabular}{|c|c|c|c|} \hline
Quantity& This paper  & Skyrme model \cite{bek88} &NQM\\ 
\hline 
$\Delta u$ & $4\Delta_3/5-a$ &$4\Delta_3/7$ & 4/3 \\
$\Delta d$ & $-\Delta_3/5-a$ & $-3\Delta_3/7$ & $-{1/3}$ \\  
$\Delta s$ & ${\Delta}/2-a$ &$-\Delta_3/7$&0\\
$\Delta\Sigma$ & $3\Delta_3/5+\Delta/2-3a$ & 0 & 1 \\
\hline
\end{tabular}
\end{center}
\end{table}

\begin{table}[ht]
\begin{center}
\caption{Quark flavor observables (the values with * are inputs)}
\begin{tabular}{|c|c|c|c|c|c|c|} \hline
Quantity & Data   & $a=0.153$ &  $a=0.144$ &$a=0.155$& $a=0.136$&
$a=0.10$\\ 
 & &(K,$\eta$,$\eta'$)& (K,$\eta$,$\eta'$)& (K,$\eta$,$\eta'$)& 
(K only)  & $\zeta=-0.429$ \\
   & &    &        &       &\cite{song9603-smw} &SU(3)\cite{cl}  \\ 
\hline 
 $\bar d-\bar u$ & $0.147\pm 0.024$ & $0.147^*$ & $0.164^*$ & $0.130^*$&
0.147$^*$& 0.095\\
${{\bar u}/{\bar d}}$ & $[{{\bar u(x)}\over {\bar d(x)}}]_{x=0.18}=0.51\pm
0.06$& 0.65 & 0.60& 0.69&0.61 &0.65\\
${{2\bar s}/{(\bar u+\bar d)}}$ & ${{<2x\bar s(x)>}\over {<x(\bar
u(x)+\bar d(x))>}}=0.477\pm 0.051$& 0.69 & 0.71&0.65 &0.62&1.64\\
 ${{2\bar s}/{(u+d)}}$ & ${{<2x\bar s(x)>}\over
{<x(u(x)+d(x))>}}=0.099\pm
0.009$& 0.128 &0.124&0.127 &0.106& 0.213\\
 ${{\sum\bar q}/{\sum q}}$ & ${{\sum<x\bar
q(x)>}\over {\sum<xq(x)>}}=0.245\pm 0.005$ & 0.235 & 0.228&0.238&0.221 
&0.214\\
 $f_s$ & $0.10\pm 0.06$\cite{gls91} & 0.10 & 0.10  & 0.09&0.08& 0.16  \\
       & $0.15\pm 0.03$\cite{dll96} &  &   & &   &   \\
       & ${{<2x\bar s(x)>}\over {\sum<x(q(x)+\bar q(x))>}}
=0.076\pm 0.022$ & &   & & &   \\
 $f_3/f_8$ & $0.23\pm 0.05$\cite{cl}  & 0.21 &0.20&0.21   &0.20
&1/3\\
\hline
\end{tabular}
\end{center}
\end{table}

\begin{table}[ht]
\begin{center}
\caption{Quark spin observables (the values with * are inputs)}
\begin{tabular}{|c|c|c|c|c|c|c|} \hline
Quantity & Data  & $a=0.153$ &$a=0.144$ & $a=0.155$&$a=0.136$& $a=0.10$ \\ 
& &  (K,$\eta$,$\eta'$)&(K,$\eta$,$\eta'$)& (K,$\eta$,$\eta'$)&
 (K only)&$\zeta=-0.429$ \\
         &        &        &         &       &\cite{song9603-smw}&
SU(3)\cite{cl} \\ 
\hline 
$\Delta u$ & $0.85\pm 0.04$\cite{smc97} & 0.85 &0.86& 0.85&0.87& 0.91\\
 &$0.85\pm 0.03$\cite{ek} &  &   &    &   &   \\
$\Delta d$&$-0.41\pm$0.04\cite{smc97} &$-$0.40&$-$0.40&$-$0.41
&$-$0.39&$-0.35$\\ 
  &$-0.41\pm$0.03\cite{ek}& & &  &  &   \\  
$\Delta s$&$-0.07\pm$0.04\cite{smc97}&$-$0.07&$-$0.06&$-$0.07&$-0.05$
&$-$0.10$^*$\\
 &$-0.08\pm$0.03\cite{ek} & & &  &  & \\
$\Delta\bar u$, $\Delta\bar d$ & $-0.02\pm 0.11$\cite{smc96}&0&0&0&0 &0 \\
\hline 
$\Gamma_1^p$ & $0.127\pm 0.012$\cite{slac95}& 0.133 &0.134 &0.132&
0.138&0.145  \\
  &$0.136\pm 0.016$\cite{smc97} & 0.145 &0.146 &0.144  &0.150 &0.158  \\
$\Gamma_1^n$ & $-0.036\pm 0.007$\cite{slac97}& $-0.037$ &$-0.036$
&$-0.038$&$-0.032$& $-0.025$ \\
 &$-0.046\pm 0.021\cite{smc97}$ & $-0.040$ &$-0.039$ &$-0.041$ &$-0.035$ &
$-0.027$\\
$\Gamma_1^d$ & $0.042\pm 0.005$\cite{slac95} & 0.045 &0.046 &0.044  &
0.049&0.056 \\
 &$0.041\pm 0.007$\cite{smc97}&0.048  &0.049 &0.047  &0.052 &0.059 \\
\hline 
$\Delta_3$ &
1.2573$\pm$0.0028\cite{pdg96}&1.2573$^*$&1.2573$^*$&1.2573$^*$
&1.2573$^*$ &1.2573$^*$ \\
$\Delta_8$& 0.579$\pm$ 0.025\cite{pdg96}& 0.579$^*$&0.579$^*$ &
0.579$^*$&
0.579$^*$& 0.754\\
\hline
\end{tabular}
\end{center}
\end{table}

\vfill\eject


\vfill\eject

\begin{figure}[htb]
\epsfxsize=5.0in
\centerline{\epsfbox{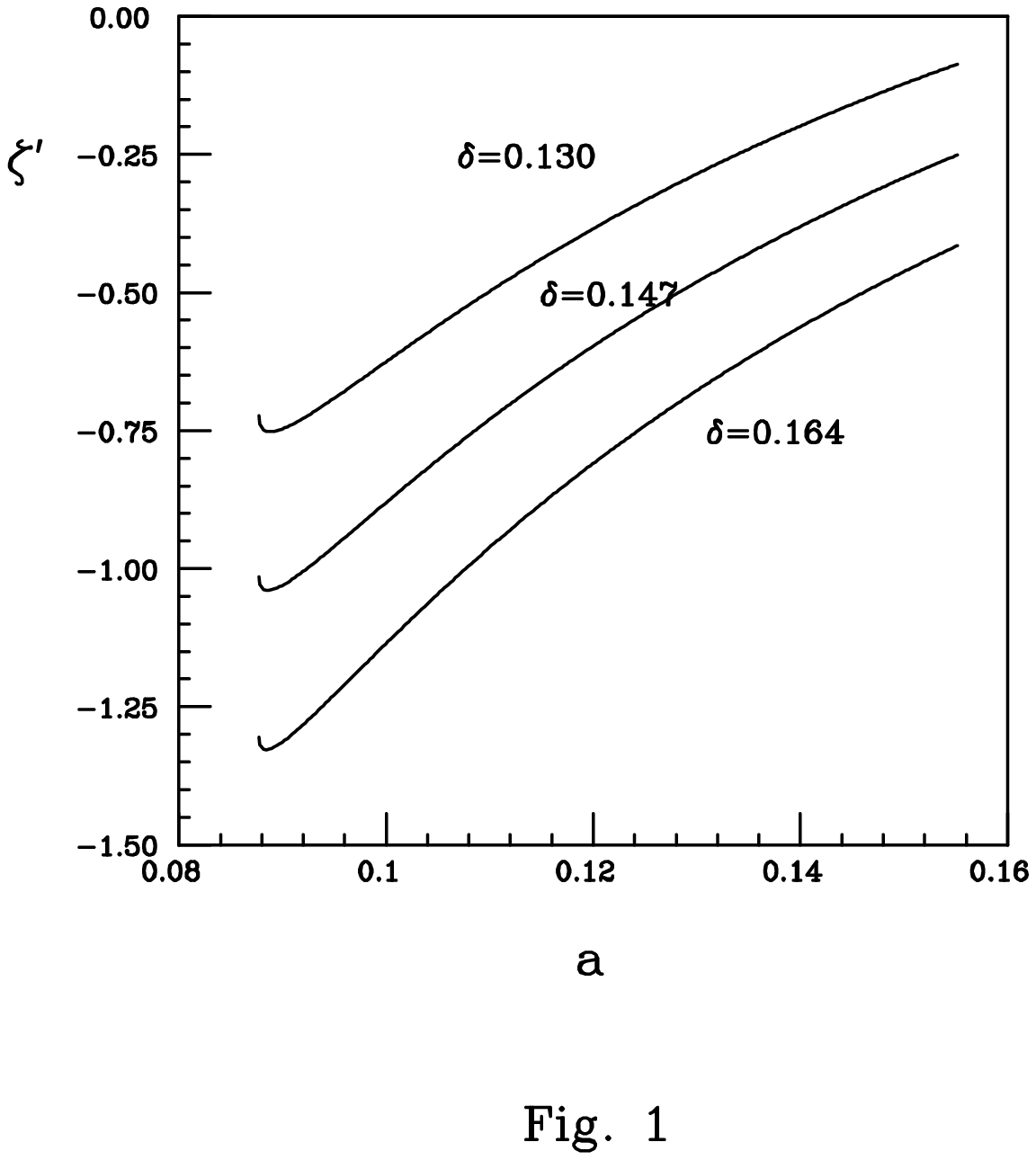}}
\caption{The U(1)-breaking parameter $\zeta'$ as function of $a$, (2.12a), 
for $\delta\equiv\bar d-\bar u$=0.130, 0.147, and 0.164.}
\end{figure}

\begin{figure}[htb]
\epsfxsize=5.0in
\centerline{\epsfbox{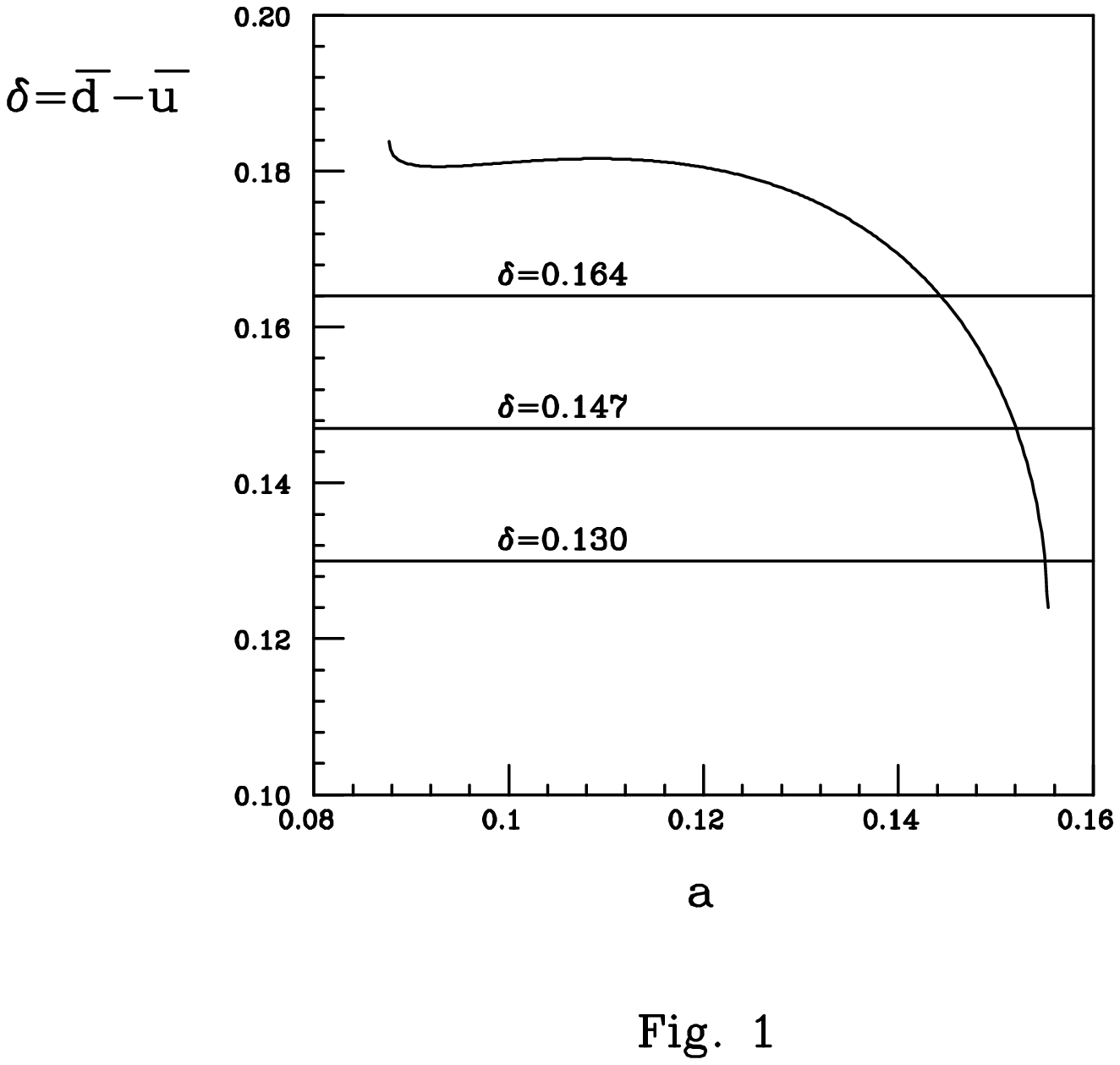}}
\caption{$\delta(a)\equiv\bar d-\bar u$ as function of $a$, from (1.5b) 
and (1.6).}
\end{figure}

\begin{figure}[htb]
\epsfxsize=5.0in
\centerline{\epsfbox{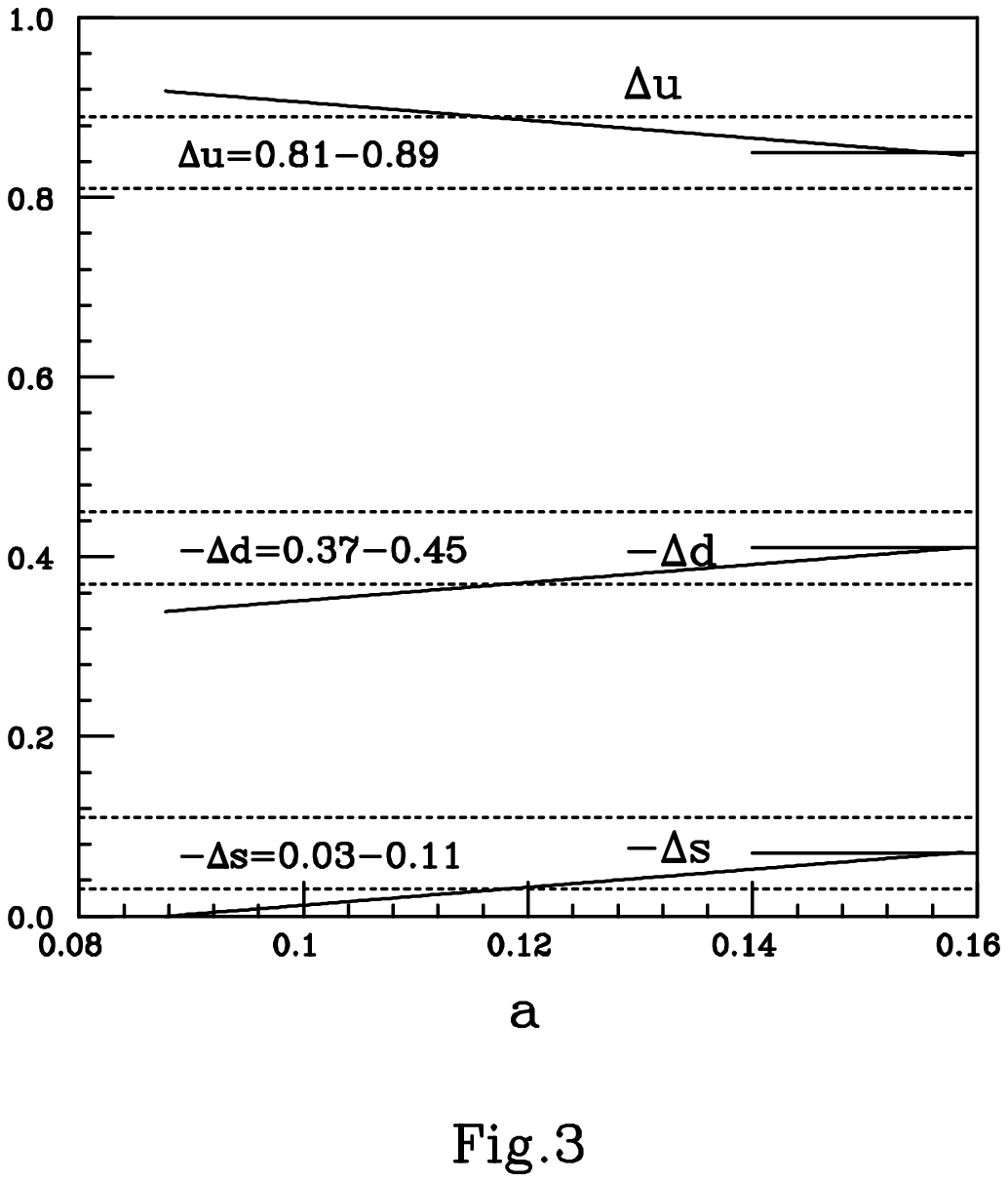}}
\caption{Quark spin contents as functions of $a$ in the symmetry 
breaking chiral quark model.}
\end{figure}

\begin{figure}[htb]
\epsfxsize=5.0in
\centerline{\epsfbox{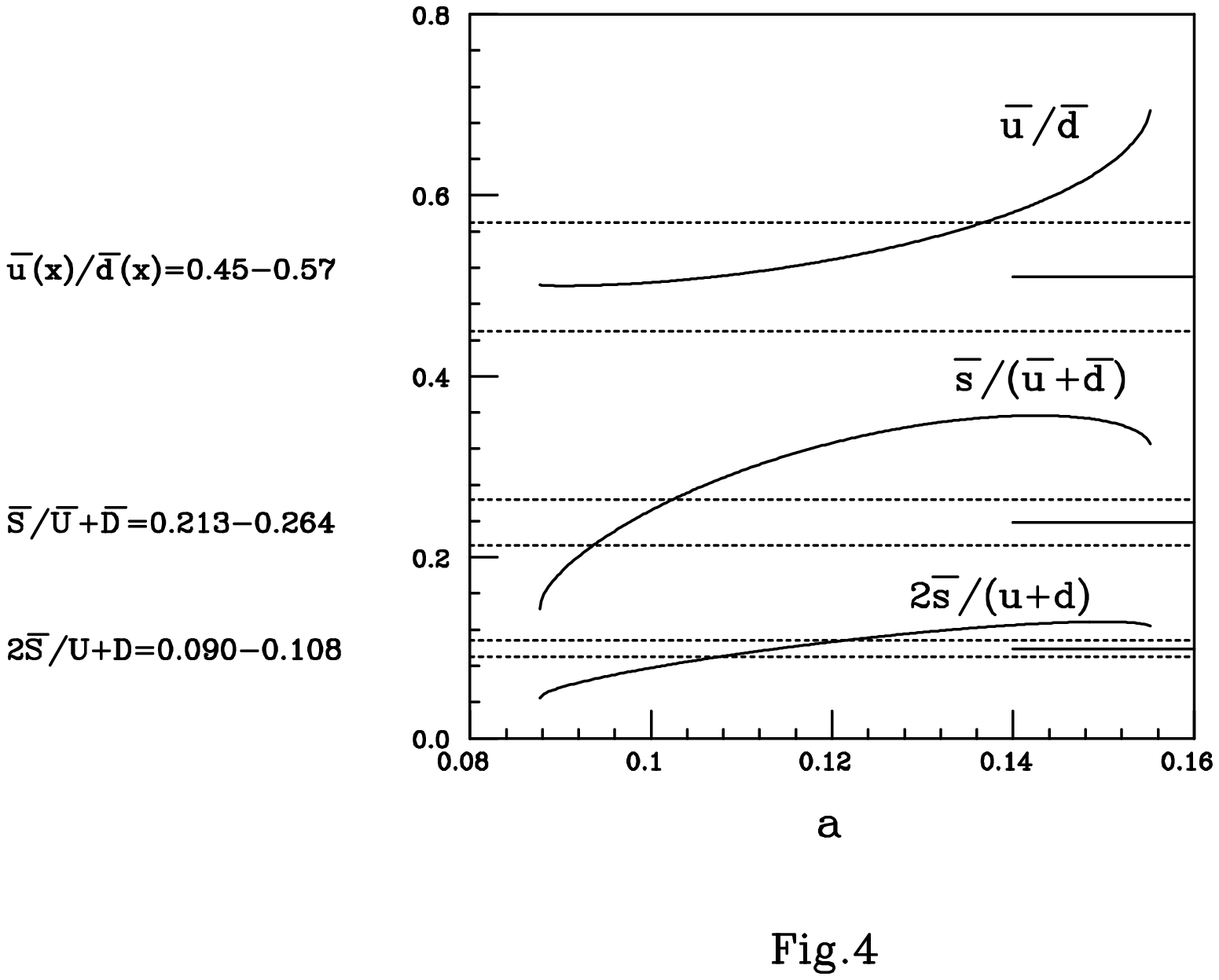}}
\caption{Quark flavor observables as functions of $a$ in the symmetry 
breaking chiral quark model. The corresponding data and comparison with
the model predictions are explained in the text.}
\end{figure}


\begin{thebibliography}{99}
\bibitem{temp} 
In the SU(3) symmetry case, we can choose $\Delta s=-0.08$ instead of 
$\Delta s=-0.10$ as input, then in the flavor sector, one has 
$\bar d-\bar u$=0.109, $\bar u/\bar d=0.55$, but 
$2\bar s/(\bar u+\bar d)=1.88$. The ratios $\bar s/(u+d)$, $\sum\bar
q/\sum q$ and $f_s$ are almost unchanged. In the spin sector, one has 
$\Delta u=0.93$, $\Delta d=-0.33$, $\Delta s=-0.08$ (input),
$\Gamma_1^p=0.158$, $\Gamma_1^n=-0.019$, $\Gamma_1^d=0.061$. 
There are no changes for the SU(3) values of $f_3/f_8$ and
$\Delta_3/\Delta_8$, hence $\Delta_8=0.754$.
\end{thebibliography}
\end{document}